\documentclass[prl,amsmath,twocolumn,showkeys,showpacs,superscriptaddress]{revtex4}
\usepackage{gensymb}
\usepackage{graphicx,color,xcolor}
\usepackage{amssymb}
\usepackage{footnote}
\usepackage{enumerate}
\usepackage{verbatim}
\usepackage{natbib}
\usepackage{float}
\usepackage[outdir=./]{epstopdf}
\usepackage{soul}

\begin{document}
  \newcommand {\nc} {\newcommand}
  \nc {\Sec} [1] {Sec.~\ref{#1}}
  \nc {\IR} [1] {\textcolor{red}{#1}} 
  \nc {\IB} [1] {\textcolor{blue}{#1}} 
  \nc {\IG} [1] {\textcolor{green}{#1}}
  \nc {\LN} [1] {\textcolor{purple}{#1}}

\title{Direct comparison between Bayesian and frequentist uncertainty quantification for nuclear reactions}

\author{G.~B.~King}
\affiliation{National Superconducting Cyclotron Laboratory, Michigan State University, East Lansing, MI 48824}
\affiliation{Department of Physics and Astronomy, Michigan State University, East Lansing, MI 48824-1321}
\author{A.~E.~Lovell} 
\affiliation{Theoretical Division, Los Alamos National Laboratory, Los Alamos, NM 87545}
\affiliation{Center for Nonlinear Studies, Los Alamos National Laboratory, Los Alamos, NM 87545}
\author{L.~Neufcourt}
\affiliation{National Superconducting Cyclotron Laboratory, Michigan State University, East Lansing, MI 48824}
\affiliation{Department of Physics and Astronomy, Michigan State University, East Lansing, MI 48824-1321}
\author{F.~M.~Nunes}
\email{nunes@nscl.msu.edu}
\affiliation{National Superconducting Cyclotron Laboratory, Michigan State University, East Lansing, MI 48824}
\affiliation{Department of Physics and Astronomy, Michigan State University, East Lansing, MI 48824-1321}

\date{\today}


\begin{abstract}
Until recently, uncertainty quantification in low energy nuclear theory was typically performed using frequentist approaches. However in the last few years, the field has shifted toward Bayesian statistics for evaluating confidence intervals. Although there are statistical arguments to prefer the Bayesian approach, no direct comparison is available.
In this work, we compare, directly and systematically,  the frequentist and Bayesian approaches to quantifying uncertainties in direct nuclear reactions. 
Starting from identical initial assumptions, we determine confidence intervals associated with the elastic and the transfer process for both methods, which are evaluated against  data via a comparison of the empirical coverage probabilities. 
Expectedly, the frequentist approach is not as flexible as the Bayesian approach in exploring parameter space and often ends up in a different minimum.  
We also show that the two methods produce significantly different correlations. 
In the end, the frequentist approach produces significantly narrower uncertainties on the considered observables than the Bayesian. 
Our study demonstrates that the uncertainties on the reaction observables considered here within the Bayesian approach  represent reality more accurately than the much narrower uncertainties obtained using the standard frequentist approach.
\end{abstract}


\maketitle

{\it Introduction:}
\label{intro}
Over the past few years, Bayesian methods have rapidly drawn much attention in the field of low energy nuclear physics. They have opened new doors for theoretical predictions:  providing a means to rigorously quantify uncertainties and the potential to help plan for future experiments at rare isotope facilities worldwide. 
Bayes statistics has been used in studies of the nuclear force \cite{furnstahl2015,melendez2017}, of nuclear stability \cite{neufcourt2018}, and in nuclear astrophysics simulations  (e.g. \cite{orford2018,yeunhwan2018}).
Our group has also been exploring Bayesian methods in the context of nuclear reactions \cite{Lovell2018}, particularly in connection to assessing the uncertainties in the predicted cross section coming from the optical potentials. While the explosion of applications of Bayesian statistics to low energy nuclear theory is very exciting, it also calls for special scrutiny. 

Up to a decade ago, there were very few studies done on uncertainty quantification in low energy nuclear theory, 
and those efforts relied primarily on $\chi^2$-minimization techniques stepping through  parameter space in the maximal direction of the local gradient. 
Relevant to our subfield, the major global optical potentials being used currently in reaction calculations \cite{bg69,ch89,kd2003} have been optimized in this way, which from now on we shall refer to as the {\it frequentist approach}.  One important conclusion from this approach was the presence of strong correlations between some parameters of the optical potential, as discussed in \cite{Lovell2017,King2018}.

In contrast,  the Bayesian framework often relies on  Markov Chain Monte Carlo (MCMC) methods to sample parameter space and obtain posterior predictions from the product of the likelihood function and prior distribution, without any normality assumption on the distribution of parameters. 
Thus, one should not assume that the minimum obtained from the frequentist approach matches the results obtained in the Bayes-MCMC method, and, moreover, that the uncertainties estimated by both methods are consistent. It is also possible that the insight obtained in the Bayesian analysis does not match the knowledge that we have established over the years using the frequentist methods. \footnote{As a disclaimer, what we here call the {\it frequentist or Bayesian approaches} corresponds to a  subset of a larger array of methods available under these umbrellas.}

For all these reasons, it is important and timely to directly compare the two approaches. Despite the many recent applications mentioned above, to our knowledge, there exists no systematic and controlled comparison between these two methodologies in our field. This is the goal of the present study.


{\it Methods and numerical details:}
\label{method}
In this study, we focus on deuteron induced reactions on heavy ions, within a three-body model \cite{ReactionsBook}. The inputs to the cross sections calculated are nucleon-nucleus optical potentials $U_{opt}$. To capture the complex many-body effects of nucleon-nucleus scattering, the potential  contains both a real part and an imaginary part to account for flux that leaves the elastic channel. Typically, optical potentials contain i) a real Woods-Saxon volume term with parameters $V$, $r$, $a$ for the depth, radius, and diffuseness ii) an imaginary Woods-Saxon volume term with parameters $W$, $r_w$, $a_w$, iii) a surface (derivative of a Woods-Saxon) imaginary term with parameters $W_s$, $r_s$, and $a_s$, iv) a spin-orbit term, and v) a Coulomb term for charged projectiles \cite{ReactionsBook}. In this work, we fitted the real volume, imaginary volume, and imaginary surface parameters and kept the spin-orbit and Coulomb terms fixed. We then used elastic scattering angular distribution data to constrain these free parameters.

We first consider neutron and proton scattering  on $^{48}$Ca, $^{90}$Zr and $^{208}$Pb. Table \ref{tab:data} (columns 1-3) contains a summary of all the data used to constrain the necessary optical potentials. 
We chose only data sets with a wide angular distribution and small error bars,
and assign a 10\% uncertainty on all data to account for experimental error and model inadequacy; this also prevents overfitting. Because we eventually propagate the uncertainties to the transfer (d,p) channel, the energies for the neutron and proton elastic scattering are carefully chosen to be close to  either half the deuteron beam energy or the energy of the proton in the exit channel.  Starting from the original Becchetti and Greenlees (BG) global parameterization for the optical potential \cite{bg69}, we use both, the frequentist and Bayesian method, to obtain posterior distributions for the parameters and 95\% confidence intervals for elastic scattering data.  In the case of $^{90}$Zr, the  resulting geometry parameters (radius and diffuseness) for  proton and neutron optical potentials resulting from the frequentist approach were significantly different from each other, when starting from the BG parameterization. Since this is physically implausible, we use the parameters from  the fitting of $^{90}$Zr(n,n) to initialize the $^{90}$Zr(p,p) reaction, only.

\begin{table}[h]
\begin{center}
\begin{tabular}{|c|c|c| r | r |}
\hline Reaction &Energy (MeV) &Ref. & \%Data(F)  & \%Data(B)  \\ \hline
$^{48}$Ca(p,p) &14.03 &\cite{48cap14}&  70 & 100  \\ \hline
$^{48}$Ca(n,n) &12  &\cite{48can12}& 61  & 100 \\ \hline
$^{48}$Ca(p,p) &25 &\cite{48cap25}& 51 &  94 \\ \hline
$^{90}$Zr(p,p) &12.7 &\cite{90zrp12}& 66 &  100 \\ \hline
$^{90}$Zr(n,n) &10 &\cite{90zrn10}&  92 &  100 \\ \hline
$^{90}$Zr(p,p) &22.5 &\cite{90zrp22}&  67 &  92 \\ \hline
$^{208}$Pb(p,p) &16 &\cite{208pbp16}&  62 &  92\\ \hline
$^{208}$Pb(n,n) &16.9 &\cite{208pbn16}&  62 &  76 \\ \hline
$^{208}$Pb(p,p) &35 &\cite{208pbp35}& 44 &  88 \\ \hline
\end{tabular}
\end{center}
\caption{List of reactions studied:  beam energy (col. 2),  data reference (col. 3), percentage of the data that falls into the 95\% confidence interval obtained in the frequentist approach (col. 4) and the Bayesian approach (col. 5). }
\label{tab:data}
\end{table}

The frequentist approach follows the methods described in detail in \cite{Lovell2017,King2018}, considering only the  uncorrelated $\chi^2$ function. 
To initialize the fitting procedure, the original BG parameters were used and allowed to vary within a wide window (physical limits for each parameter). Then $1600$ parameter sets were pulled from the multivariate normal distribution induced by the $\chi^2$ around the optimal parameters \cite[Eq. (6)]{Lovell2017}; 95\% confidence intervals were constructed by removing the highest and lowest 2.5\% of the predicted values for the cross sections at each calculated angle.

\begin{figure*}
	\includegraphics[width=.32\linewidth]{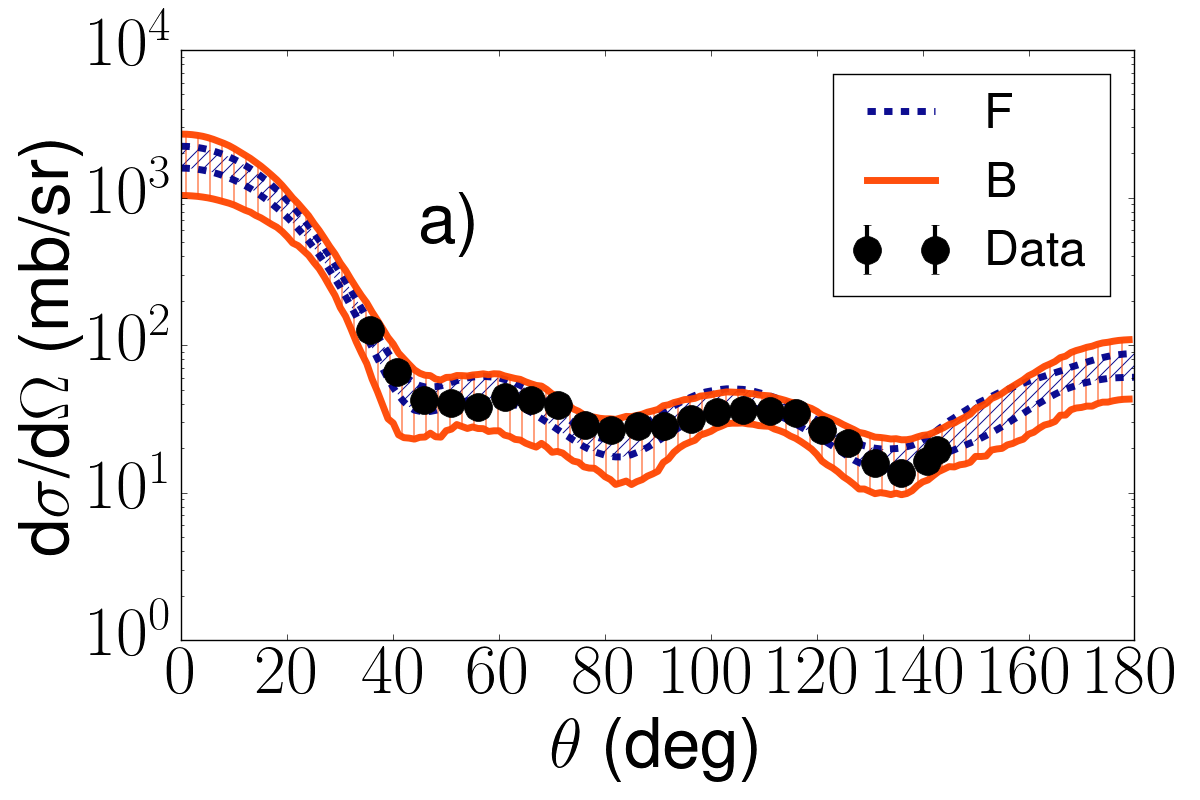}
	\includegraphics[width=.32\linewidth]{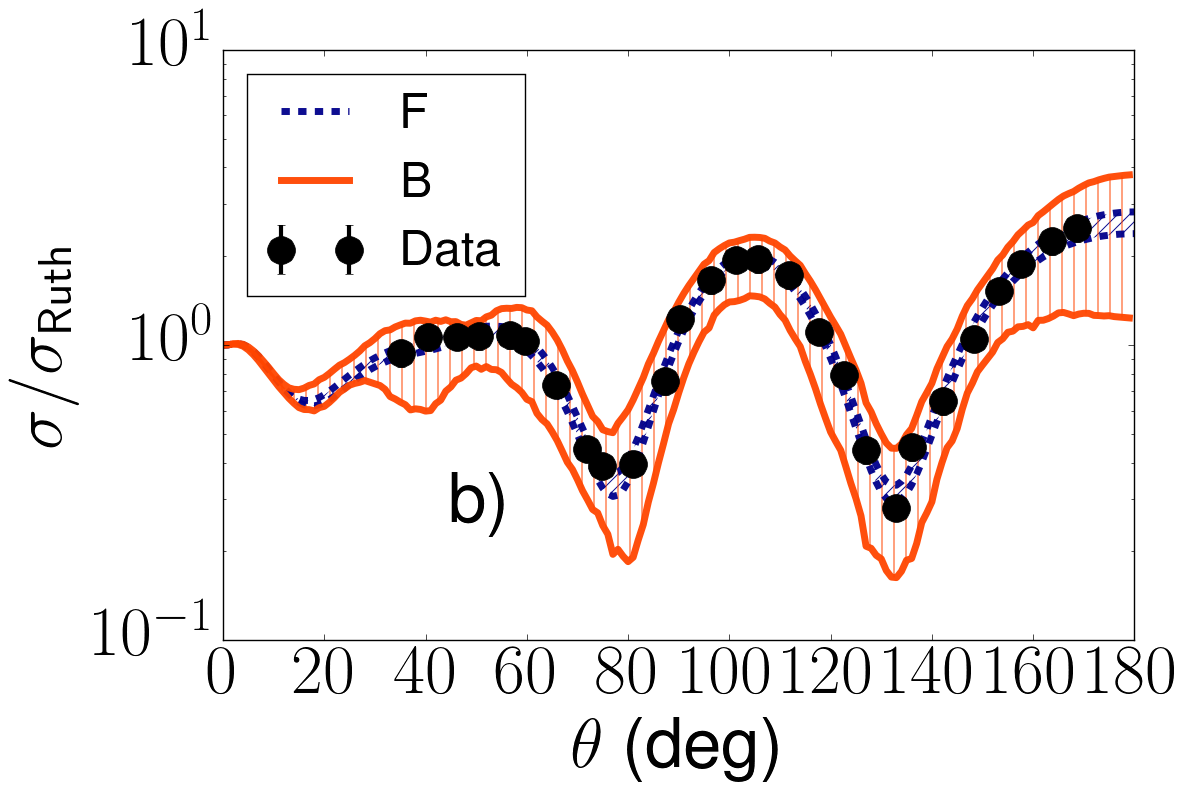}
	\includegraphics[width=.32\linewidth]{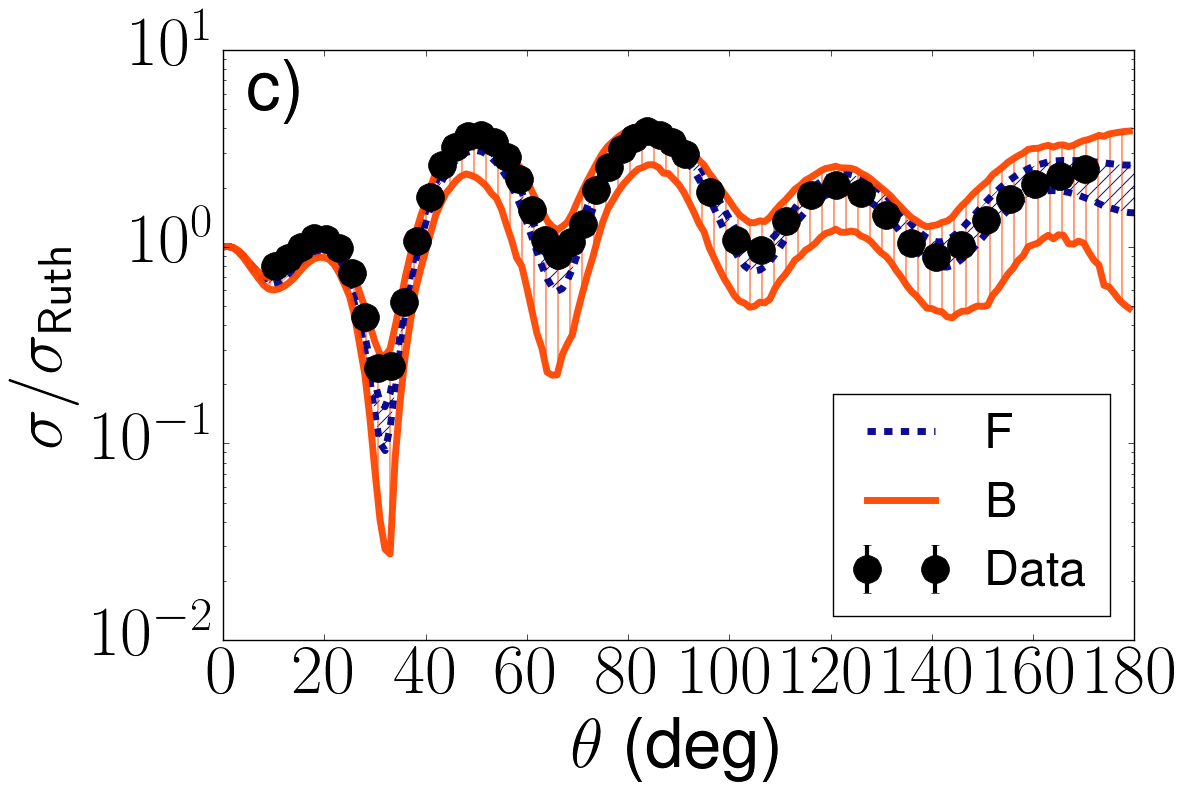} \\
	\includegraphics[width=.32\linewidth]{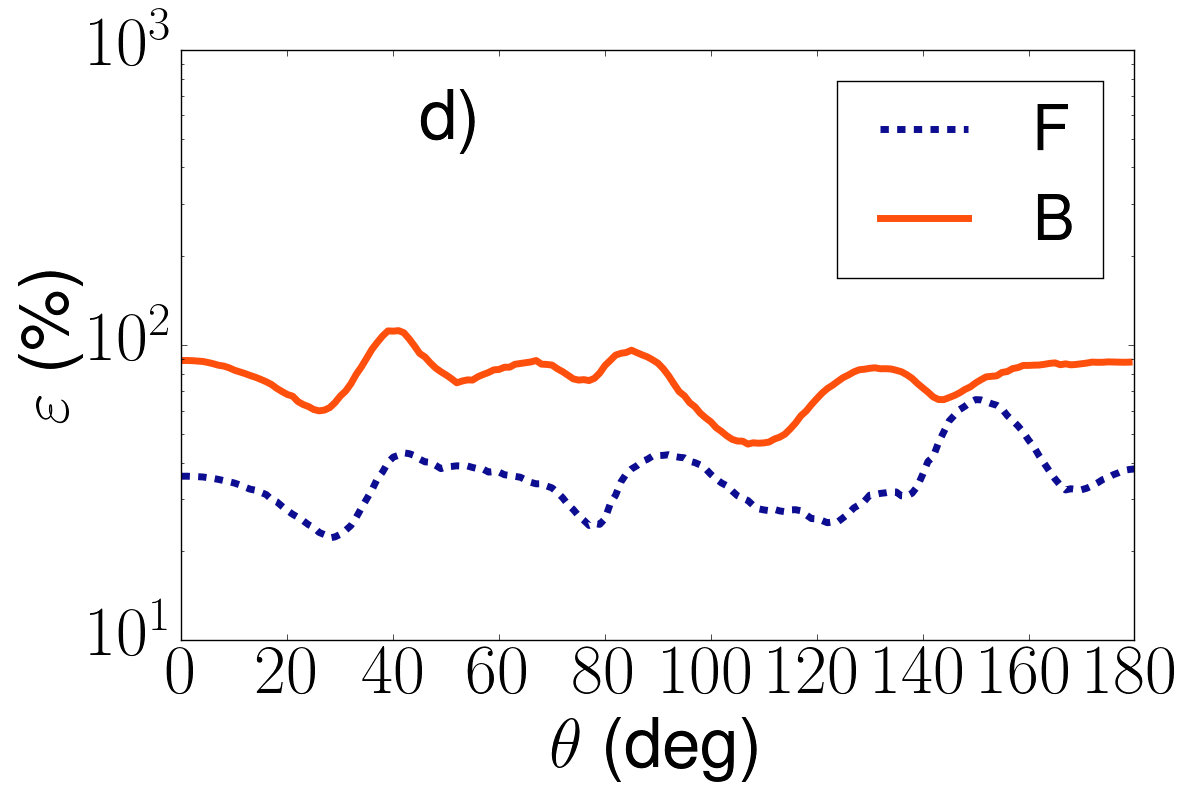}
	\includegraphics[width=.32\linewidth]{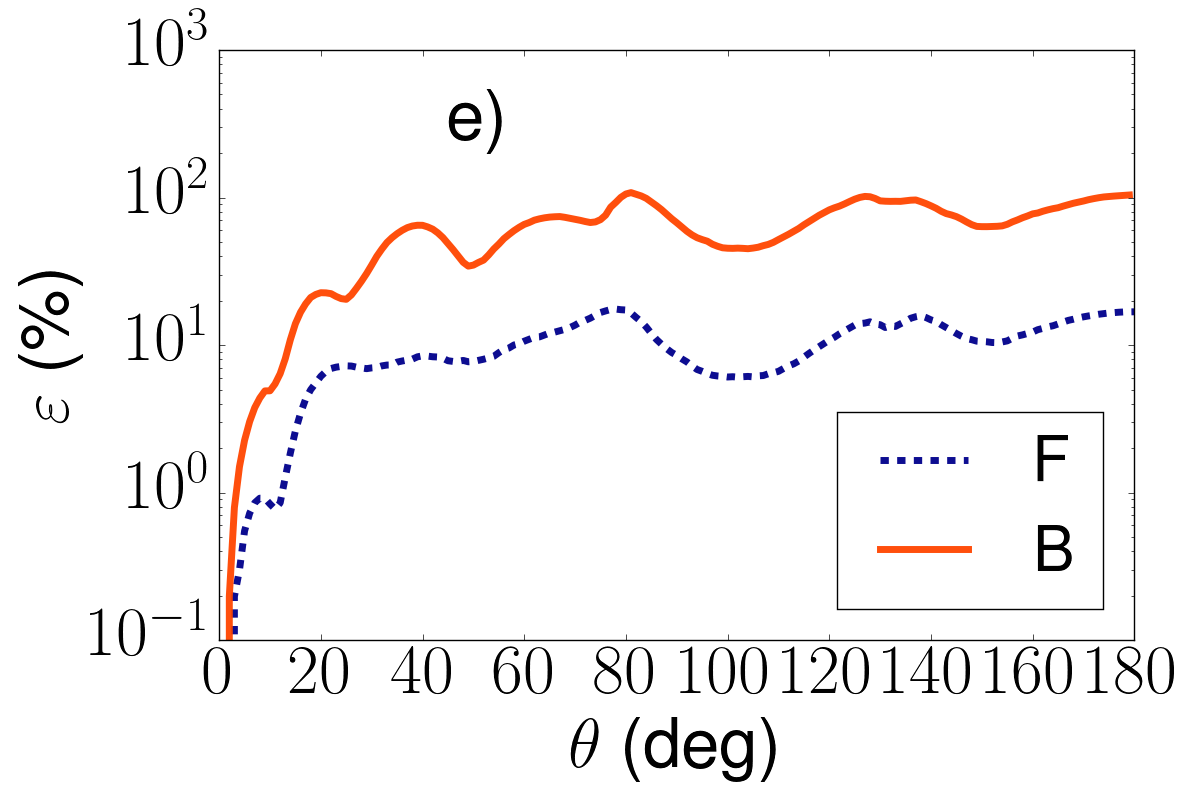}
	\includegraphics[width=.32\linewidth]{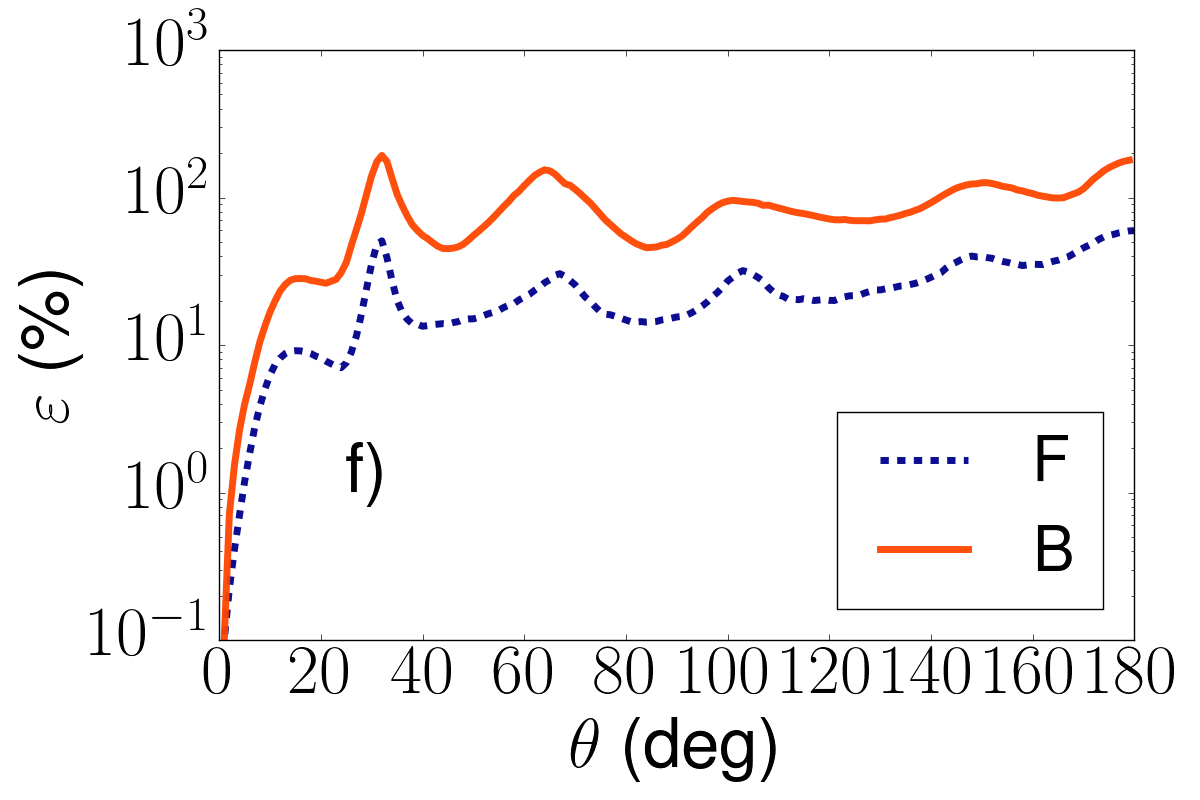} \\
	\includegraphics[width=.32\linewidth]{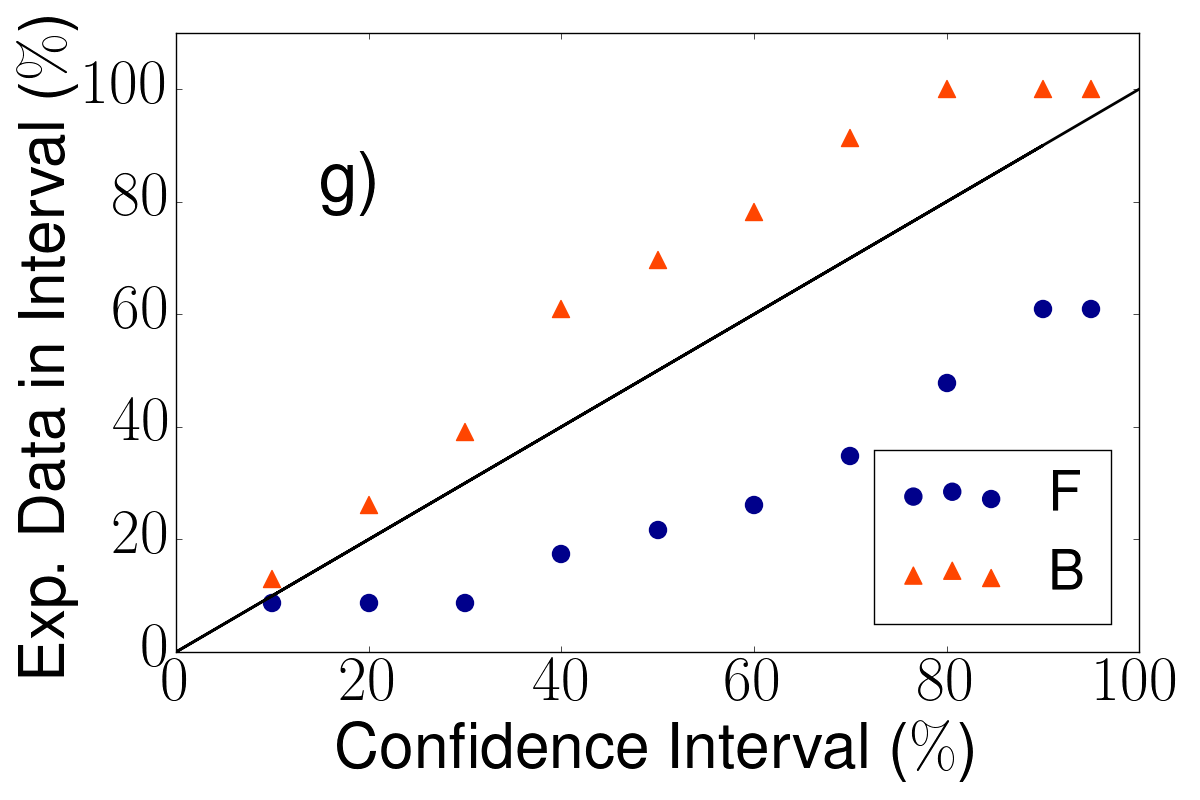}
	\includegraphics[width=.32\linewidth]{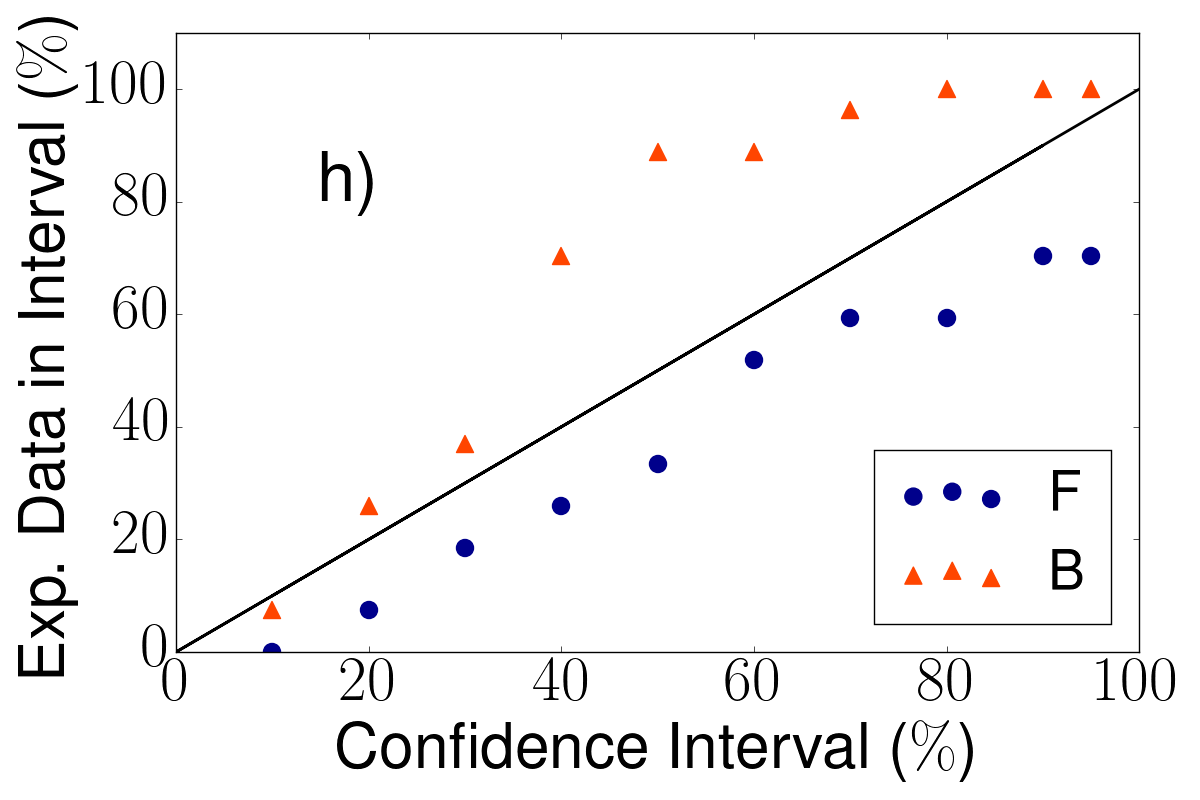}
	\includegraphics[width=.32\linewidth]{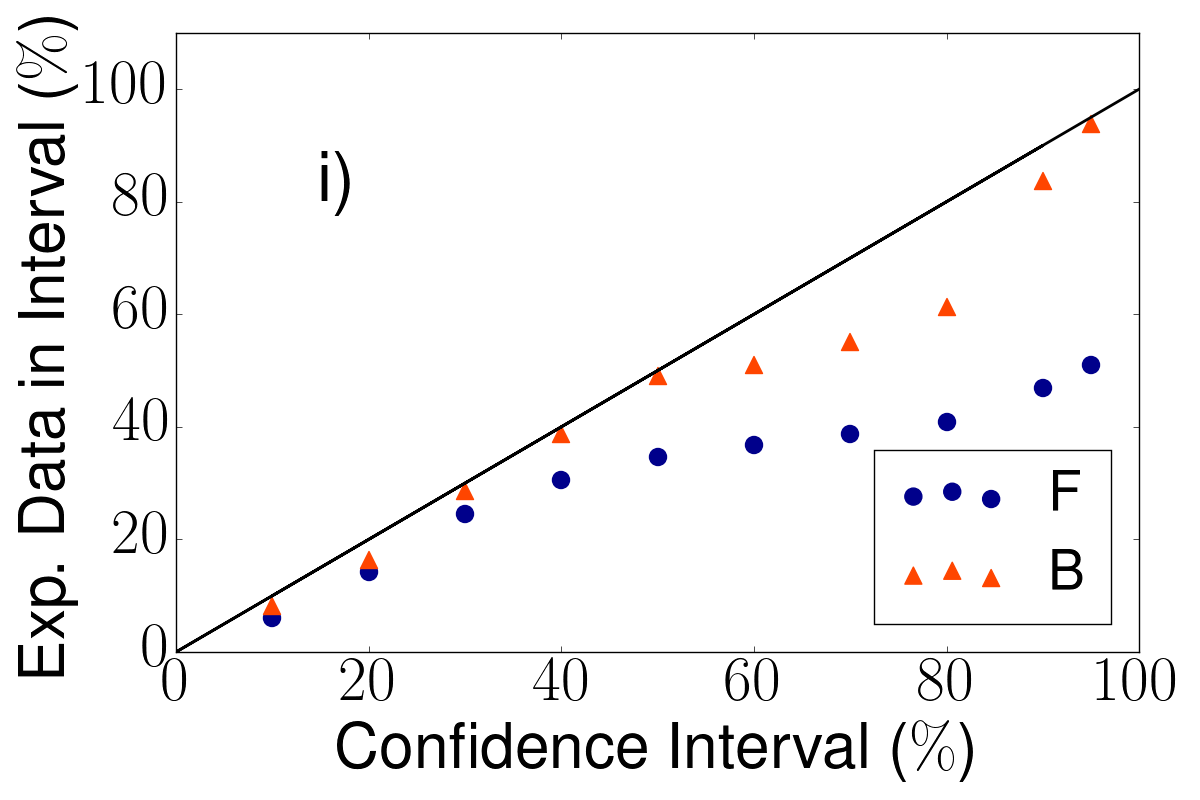} 
\caption{Elastic scattering for neutrons and protons on $^{48}$Ca. Top row (panels a,b,c): the predicted $95$\% confidence intervals from Bayesian (orange vertical hash) is compared  with the frequentist approach (blue slanted hash), and to experimental data.  Middle row (panels d,e,f):  percent uncertainty of the confidence intervals.  Bottom row (panels g,h,i):  comparison of the percentage of data falling within the given confidence interval. First column: $^{48}$Ca(n,n) at 12 MeV; Second column: $^{48}$Ca(p,p) at 14 MeV; Third column: $^{48}$Ca(p,p) at 25 MeV.}
	\label{fig:elastic} 
\end{figure*}

We use the Bayesian MCMC implementation  \cite{Lovell2018}, with the same numerical details.
We choose wide Gaussian priors, centered at the original BG value, with a standard deviation equal to the mean value of the distribution. This choice has proven to ensure that the parameter space is appropriately sampled \cite{Lovell2018}. Again a total of 1600 parameter sets were collected.  In contrast to the frequentist method, 95\% confidence intervals are constructed by taking the densest 95\% of the 1600 cross section values calculated at each angle.
 The wrapper codes developed for both the frequentist  and the Bayesian analyses make use of the reaction codes {\sc fresco} and {\sc sfresco}  \cite{fresco}.

{\it Results:} 
\label{results}
We illustrate our results in detail for reactions on the $^{48}$Ca target. We start by analyzing nucleon elastic scattering on $^{48}$Ca in Fig.\ref{fig:elastic}. In the first row, we present  95\% confidence intervals for neutron and proton elastic-scattering angular distributions as a function of scattering angle.
For both methods, the uncertainty intervals appear to provide a good description of the data. However, the Bayesian approach results in wider confidence intervals than the frequentist. This is clear in Figs.\ref{fig:elastic}(panel d, e and f), where the percentage width of each uncertainty band is computed as a function of scattering angle. Here, $\epsilon$ was obtained by taking the width of the confidence interval at a given angle in the distribution and dividing by the best fit (average) cross section value in the frequentist (Bayesian) approach.

\begin{figure}[t!]
	\includegraphics[width=.93\linewidth]{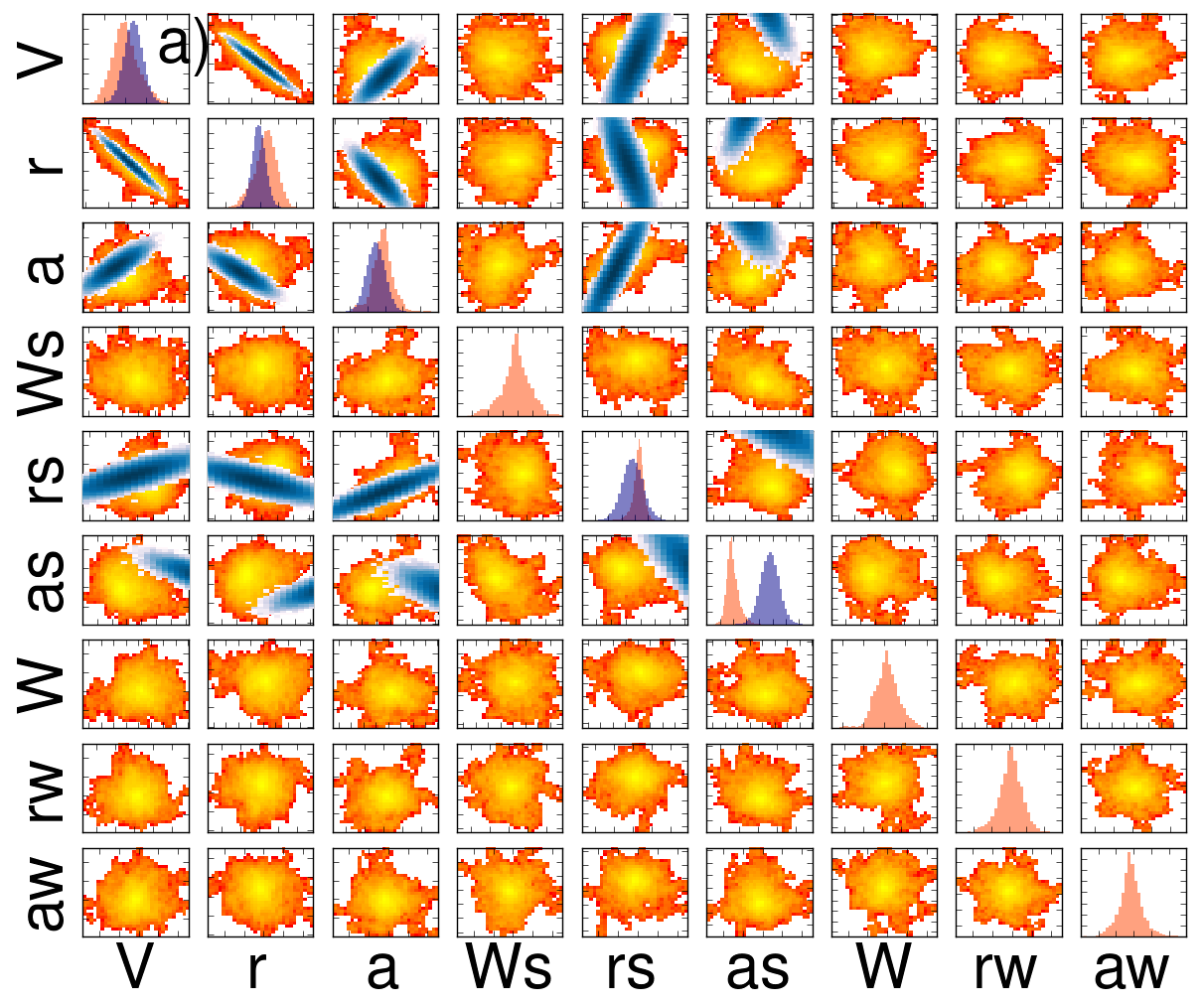}\\
	\includegraphics[width=.93\linewidth]{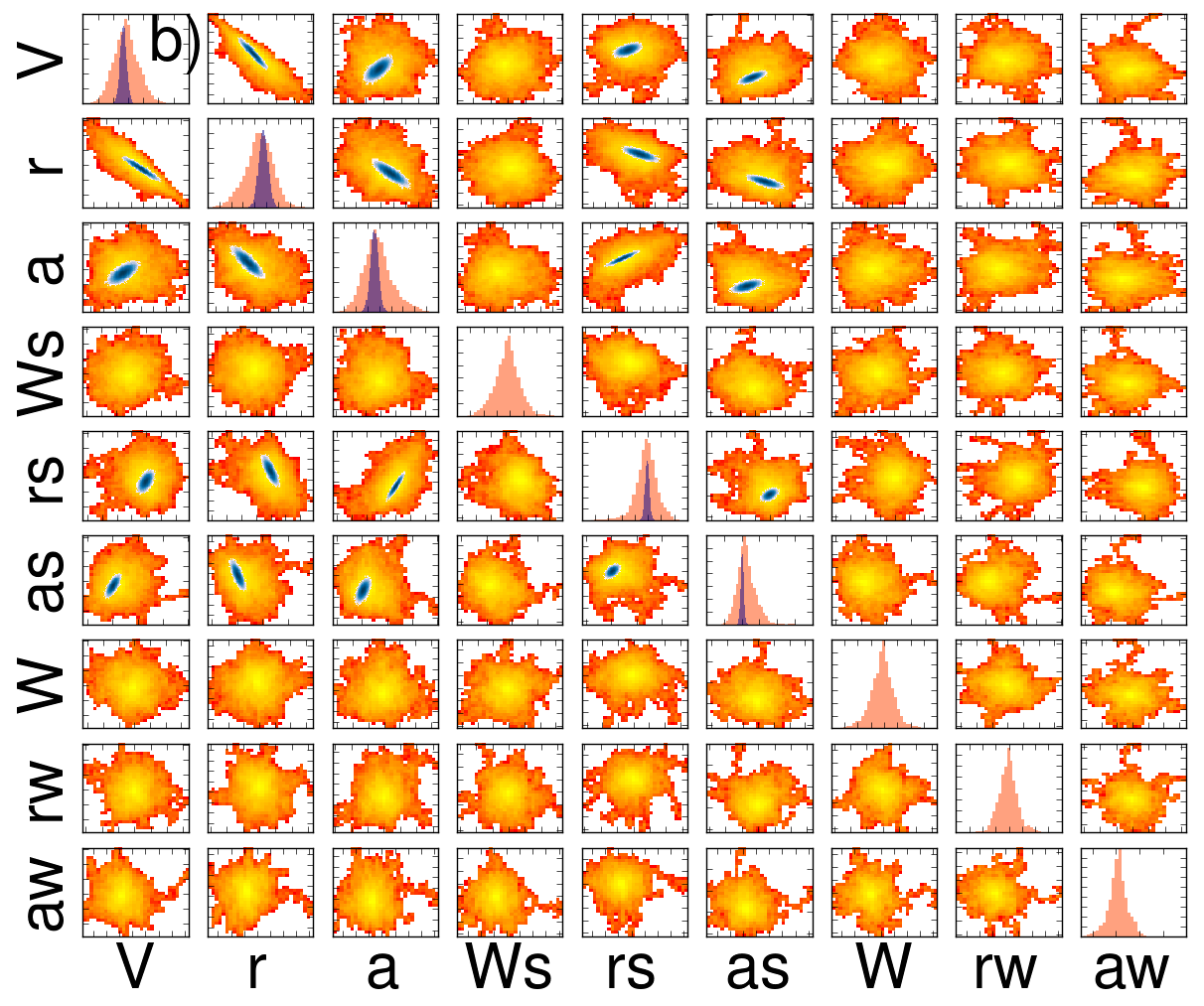}\\
	\includegraphics[width=.93\linewidth]{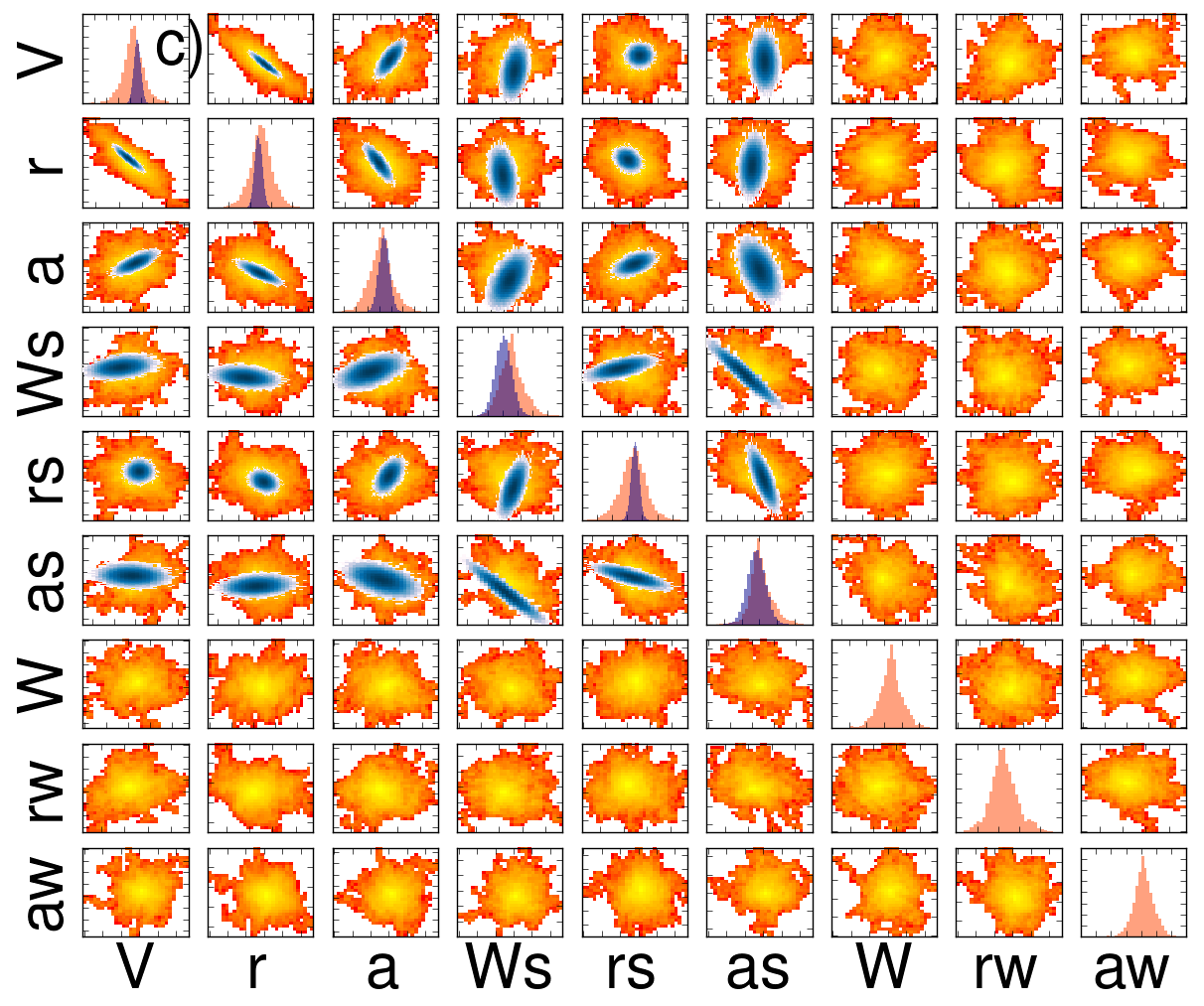}
\caption{Posterior distributions for the parameters (diagonal) and scatter plots for the correlations between parameters: Bayesian are shown in shades of orange and frequentist in shades of blue: $^{48}$Ca(n,n) at 12 MeV (a); $^{48}$Ca(p,p) at 14 MeV (b); and $^{48}$Ca(p,p) at 25 MeV (c). Depths are in MeV and radii and difuseness are in fm. }
	\label{fig:scatter} 
\end{figure}

The reliability of the confidence intervals obtained from the two statistical methods can be assessed by the empirical coverage of the confidence  intervals. We show in  Figs.\ref{fig:elastic} (g)(h) and (i) the empirical coverage probability curves of each model, namely the percentage of the experimental data which actually falls within the predicted  confidence intervals with nominal value spanning $[0, 100\%]$. Because this corresponds to training data, one might expect that reality matches predictions; this ideal situation corresponds to the black diagonal line in Figs.\ref{fig:elastic}(h, i, j).  Points below this line correspond to underestimation of the confidence intervals, whereas points above the diagonal correspond to confidence intervals that are too large (a lesser harm).
For each of the three reactions shown in Fig.\ref{fig:elastic}, the Bayesian model provides an accurate quantification of its uncertainty for the larger confidence intervals. In contrast, the frequentist approach undershoots when large confidence intervals are considered. This indicates that the 95\% confidence intervals obtained with the frequentist approach are unrealistically narrow.

We now inspect the posterior distributions for the parameters and the correlations between parameters. In Fig.\ref{fig:scatter}, we show, along the diagonal, the posterior distributions predicted by both approaches (blue for the frequentist and orange for the Bayesian): $^{48}$Ca(n,n) at 12 MeV in Fig.\ref{fig:scatter}(a); $^{48}$Ca(p,p) at 14 MeV in Fig.\ref{fig:scatter}(b); and $^{48}$Ca(p,p) at 25 MeV in Fig.\ref{fig:scatter}(c).
Note that these correlation plots require larger statistics: for Fig.\ref{fig:scatter} we needed to collect $100,000$ parameter sets.
The peak of the posterior distributions obtained for the parameters associated with the real part of the optical potential are nearly identical in both approaches, 
however the Bayesian procedure produce densities with large and sometimes asymmetric tails which thus differ significantly from Gaussian distributions.
The differences in the peak of the posteriors for the parameters associated with the imaginary components  are more noticeable. Note that some of these imaginary parameters could not be included in the fit for the frequentist approach, as the minimization procedure drove the parameters into unphysical values. For those cases, the parameters in the frequentist model were fixed at some intermediate physically plausible values, and only the Bayesian posteriors are shown.

Fig.\ref{fig:scatter} also contains the scatter plots associated with each pair of parameters. Elongated elliptical forms demonstrate high correlations between parameter pairs, while circular scatter plots indicate low correlations. It is clear from Fig.\ref{fig:scatter} that  the frequentist approach finds strong correlations between parameters. This was also the conclusion in \cite{Lovell2017} and has been well established in the field of nuclear reactions \cite{ReactionsBook}. 
However, the Bayesian approach provides a very different picture: we find systematically that only the  depth and radius of the real part are strongly correlated, in all cases here considered. 
All other scatter plots have approximately circular shape indicating that the parameters are weakly correlated. 



The reduction of the number of fitted parameters in the frequentist approach could be responsible for the additional correlations seen in the scatter plots in the frequentist approach. To test this hypothesis, we have rerun the Bayesian procedure, fixing the same parameters as in the frequentist approach. The results lead to the same overall conclusion.
Given that the posterior distributions for the Bayesian deviate strongly from Gaussian, one can thus  expect that the bivariate distributions of the Bayesian model (corresponding to the scatter plots in Fig.\ref{fig:scatter}) would also deviate. 
Our hypothesis is thus that this is causing a bias in the frequentist approach, which leads to erroneous conclusions, both in the magnitude of the uncertainties and in the correlations between parameters.

\begin{figure}
	\includegraphics[width=.95\linewidth]{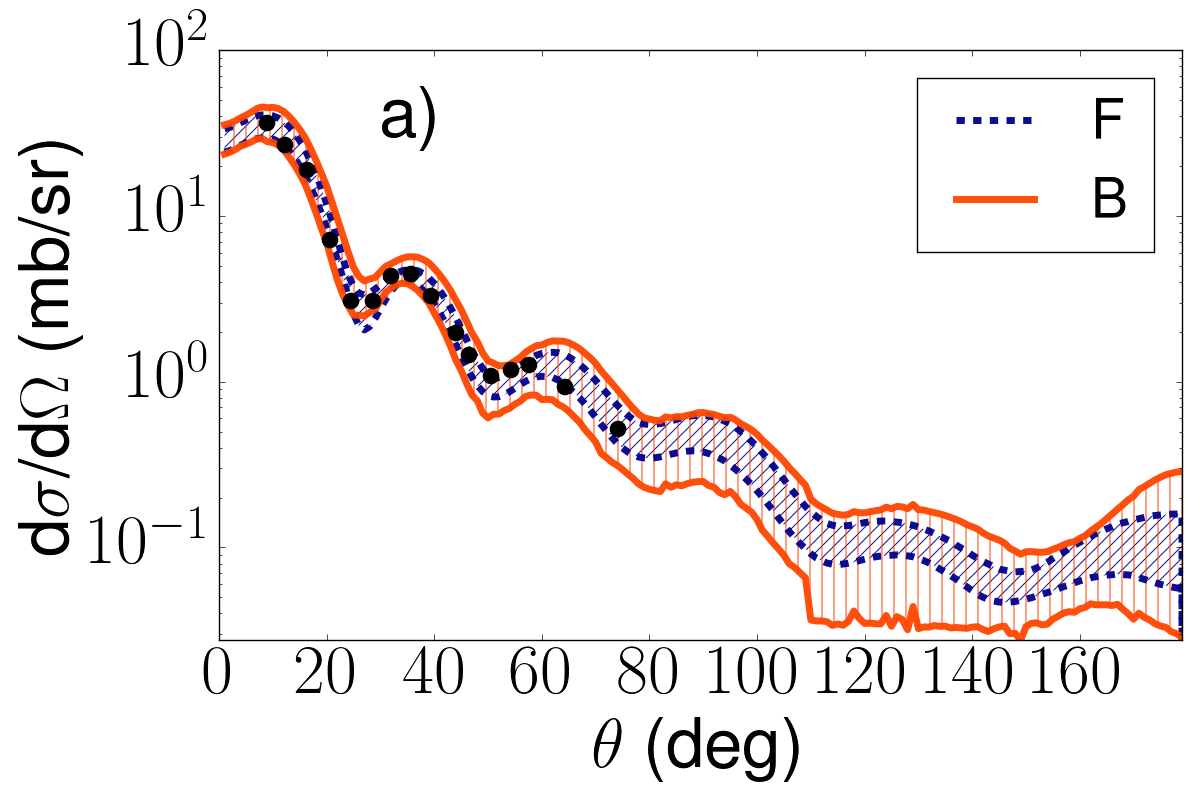}
	\includegraphics[width=.95\linewidth]{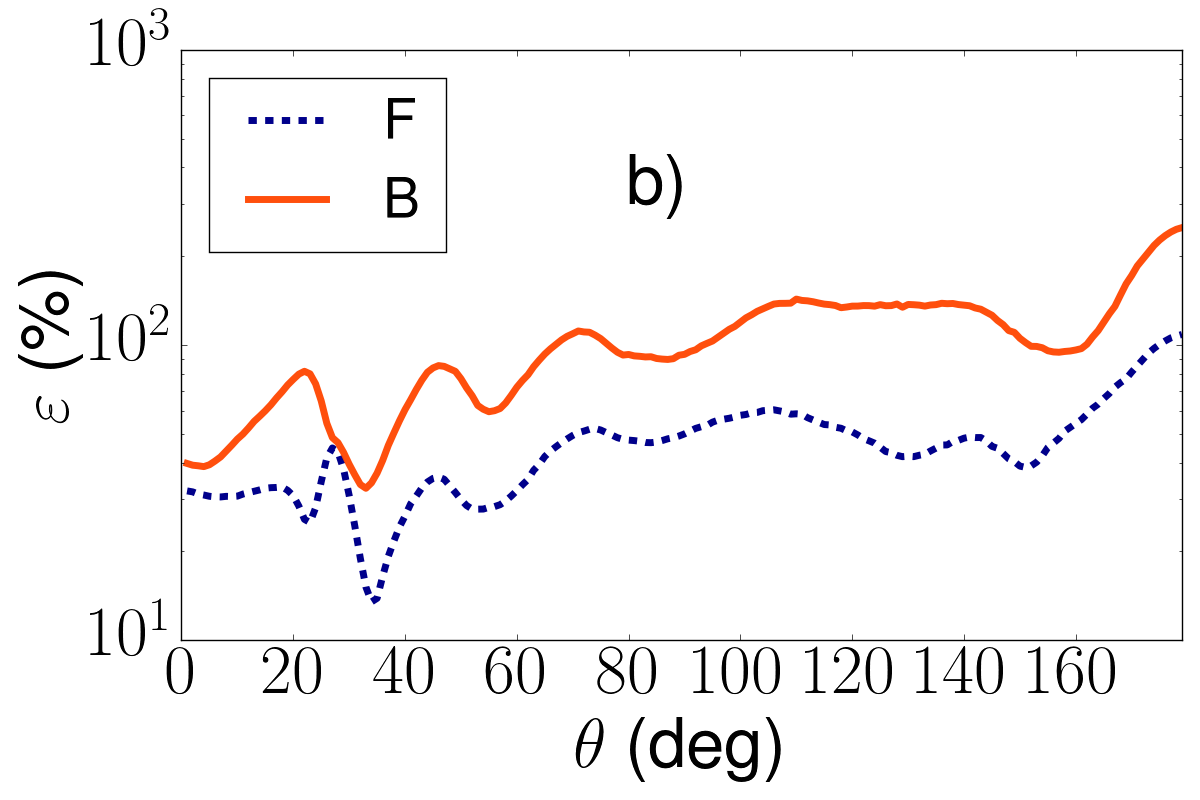}
	\includegraphics[width=.95\linewidth]{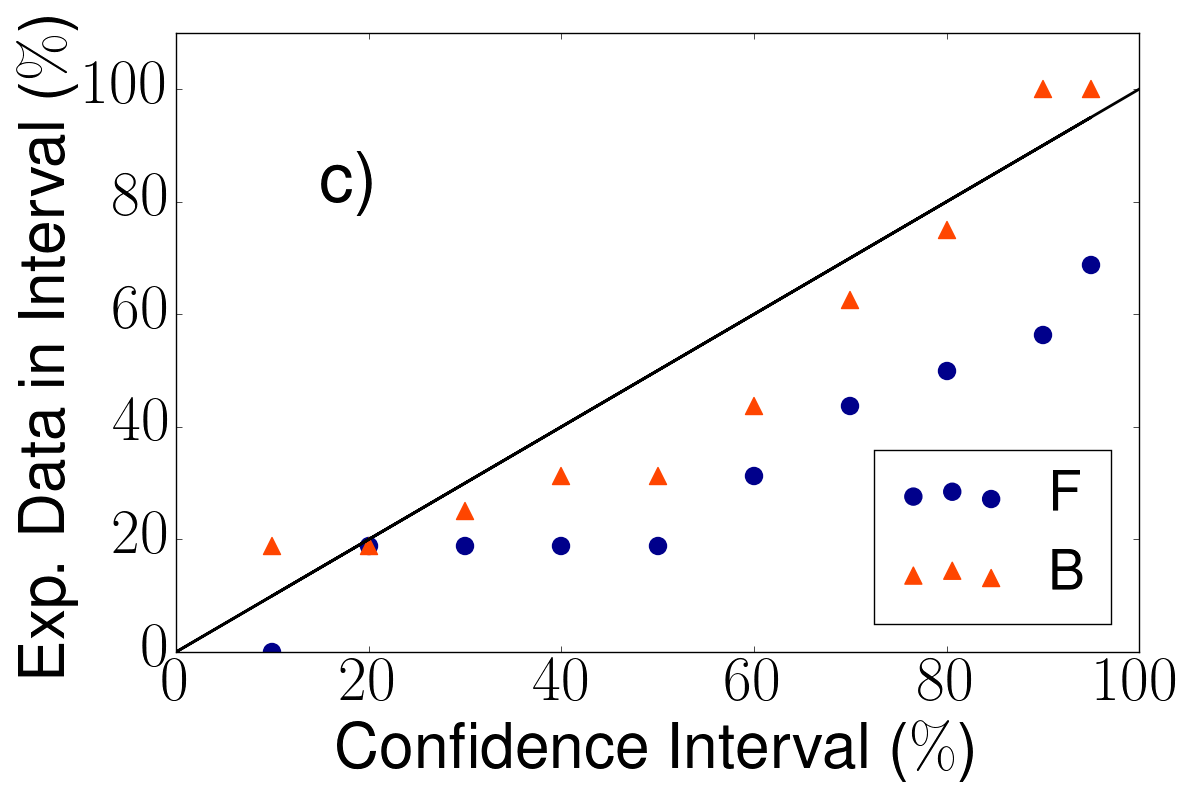}
\caption{Transfer cross sections for $^{48}$Ca(d,p): the predicted $95$\% confidence intervals from the Bayesian (orange vertical hash) and the frequentist approach (blue slashed hash) (panel a); percent uncertainty of the confidence intervals (panel b); comparison of the percentage of data falling within the given confidence interval (panel c).}
	\label{fig:transfer} 
\end{figure}

Finally, we now look at the results when propagated through a three-body reactions model to determine the one-nucleon transfer (d,p) cross sections as done in \cite{Lovell2018,King2018}. The adiabatic wave approximation (ADWA) \cite{Johnson1974} was the reaction model used to describe the transfer reaction. Apart from the optical potentials, all other inputs used in the calculations were the same as those used in \cite{Lovell2018}.  Using the parameter posteriors for each of the relevant nucleon-target optical potentials,  we obtain 95\% confidence intervals for the transfer reaction (our wrapper codes call the reaction code \textsc{NLAT} \cite{nlat} for this purpose). We show, in Fig.\ref{fig:transfer}(a), the 95\% confidence intervals for the angular distributions of $^{48}$Ca(d,p)$^{49}$Ca(g.s.) at 19.3 MeV. Again the blue-slanted-hashed area corresponds to the results using the frequentist approach and the orange-vertical-hashed area corresponds to those using the Bayesian approach. The transfer cross sections have been normalized to the data taken from Ref.\cite{ca48dp} at the peak of the angular distribution.
Fig.\ref{fig:transfer}(b) shows that the percent uncertainty of the confidence intervals in the Bayesian approach is larger than the frequentist across all angles. Finally, we also show the percentage of this test data  that falls within the given confidence interval in Fig.\ref{fig:transfer}(c). 
As previously with the training data, the Bayesian approach is more reliable, particularly when considering  higher-levels of confidence.

In order to be able to draw general conclusions we expand our study to include reactions on $^{90}$Zr and $^{208}$Pb, as listed in Table \ref{tab:data}. We include in columns 4 and 5, the percentage of data that actually falls in the 95\% confidence interval predicted by the frequentist approach and the Bayesian approach, respectively. For all but one of the cases considered, the frequentist approach does not provide a reliable estimate for the confidence interval. Indeed, the confidence intervals predicted by the frequentist approach are too narrow and the fraction of the data that actually falls in them is much smaller than the nominal 95\%. On the contrary, the Bayesian approach tends to be more conservative for  the nucleon elastic scattering on $^{48}$Ca and $^{90}$Zr at the lower energies and makes predictions that are very close to the correct values for most of the other cases.


{\it Conclusions:}
\label{conclusions}
In this work we compare directly two methods to evaluate uncertainties in reaction theory, namely the standard frequentist approach and the Bayesian framework. We perform a systematic study of nucleon scattering on three different target nuclei and use elastic scattering data to constrain the optical potential parameters. Our study is also controlled in that we make the same assumptions in both the frequentist and the Bayesian approaches. We conclude that the 95\% confidence intervals generated by the frequentist approach are unrealistically narrow, as opposed to the Bayesian approach that provide a correct assessment of the uncertainty. In addition, the frequentist approach generates strong correlations between parameters that are not seen in the Bayesian approach. We attribute these differences to a strong deviation from the $\chi^2$-based  Gaussian distribution. 
When propagating these uncertainty to  (d,p) reactions, we arrive at the same conclusions deduced from the elastic scattering analysis.


There is, a priori, no reason for the stark differences found between these two uncertainty quantification approaches to be specific to reaction theory, therefore it is desirable that other subfields (e.g. atomic and molecular physics),  perform similar studies.

\bigskip

\begin{acknowledgments}
This work was supported by the National Science
Foundation under Grant  PHY-1403906, the Stewardship Science Graduate Fellowship program under Grant No. DE-NA0002135, the
Department of Energy under Contract No. DE-FG52-
08NA28552, and under the auspices of the U.S. Department of Energy by Los Alamos National Laboratory under Contract DE-AC52-06NA25396. This work relied on iCER and the High Performance Computing Center at Michigan State University for computational resources. 
All elastic scattering data were collected from the \href{https://www-nds.iaea.org/exfor/exfor.htm}{EXFOR} database.
\end{acknowledgments}

\bibliography{bayesletter}

\begin{thebibliography}{25}
\expandafter\ifx\csname natexlab\endcsname\relax\def\natexlab#1{#1}\fi
\expandafter\ifx\csname bibnamefont\endcsname\relax
  \def\bibnamefont#1{#1}\fi
\expandafter\ifx\csname bibfnamefont\endcsname\relax
  \def\bibfnamefont#1{#1}\fi
\expandafter\ifx\csname citenamefont\endcsname\relax
  \def\citenamefont#1{#1}\fi
\expandafter\ifx\csname url\endcsname\relax
  \def\url#1{\texttt{#1}}\fi
\expandafter\ifx\csname urlprefix\endcsname\relax\def\urlprefix{URL }\fi
\providecommand{\bibinfo}[2]{#2}
\providecommand{\eprint}[2][]{\url{#2}}

\bibitem[{\citenamefont{Furnstahl et~al.}(2015)\citenamefont{Furnstahl, Klco,
  Phillips, and Wesolowski}}]{furnstahl2015}
\bibinfo{author}{\bibfnamefont{R.~J.} \bibnamefont{Furnstahl}},
  \bibinfo{author}{\bibfnamefont{N.}~\bibnamefont{Klco}},
  \bibinfo{author}{\bibfnamefont{D.~R.} \bibnamefont{Phillips}},
  \bibnamefont{and}
  \bibinfo{author}{\bibfnamefont{S.}~\bibnamefont{Wesolowski}},
  \bibinfo{journal}{Phys. Rev. C} \textbf{\bibinfo{volume}{92}},
  \bibinfo{pages}{024005} (\bibinfo{year}{2015}),
  \urlprefix\url{https://link.aps.org/doi/10.1103/PhysRevC.92.024005}.

\bibitem[{\citenamefont{Melendez et~al.}(2017)\citenamefont{Melendez,
  Wesolowski, and Furnstahl}}]{melendez2017}
\bibinfo{author}{\bibfnamefont{J.~A.} \bibnamefont{Melendez}},
  \bibinfo{author}{\bibfnamefont{S.}~\bibnamefont{Wesolowski}},
  \bibnamefont{and} \bibinfo{author}{\bibfnamefont{R.~J.}
  \bibnamefont{Furnstahl}}, \bibinfo{journal}{Phys. Rev. C}
  \textbf{\bibinfo{volume}{96}}, \bibinfo{pages}{024003}
  (\bibinfo{year}{2017}),
  \urlprefix\url{https://link.aps.org/doi/10.1103/PhysRevC.96.024003}.

\bibitem[{\citenamefont{Neufcourt et~al.}(2018)\citenamefont{Neufcourt, Cao,
  Nazarewicz, and Viens}}]{neufcourt2018}
\bibinfo{author}{\bibfnamefont{L.}~\bibnamefont{Neufcourt}},
  \bibinfo{author}{\bibfnamefont{Y.}~\bibnamefont{Cao}},
  \bibinfo{author}{\bibfnamefont{W.}~\bibnamefont{Nazarewicz}},
  \bibnamefont{and} \bibinfo{author}{\bibfnamefont{F.}~\bibnamefont{Viens}},
  \bibinfo{journal}{Phys. Rev. C} \textbf{\bibinfo{volume}{98}},
  \bibinfo{pages}{034318} (\bibinfo{year}{2018}),
  \urlprefix\url{https://link.aps.org/doi/10.1103/PhysRevC.98.034318}.

\bibitem[{\citenamefont{Orford et~al.}(2018)\citenamefont{Orford, Vassh, Clark,
  McLaughlin, Mumpower, Savard, Surman, Aprahamian, Buchinger, Burkey
  et~al.}}]{orford2018}
\bibinfo{author}{\bibfnamefont{R.}~\bibnamefont{Orford}},
  \bibinfo{author}{\bibfnamefont{N.}~\bibnamefont{Vassh}},
  \bibinfo{author}{\bibfnamefont{J.~A.} \bibnamefont{Clark}},
  \bibinfo{author}{\bibfnamefont{G.~C.} \bibnamefont{McLaughlin}},
  \bibinfo{author}{\bibfnamefont{M.~R.} \bibnamefont{Mumpower}},
  \bibinfo{author}{\bibfnamefont{G.}~\bibnamefont{Savard}},
  \bibinfo{author}{\bibfnamefont{R.}~\bibnamefont{Surman}},
  \bibinfo{author}{\bibfnamefont{A.}~\bibnamefont{Aprahamian}},
  \bibinfo{author}{\bibfnamefont{F.}~\bibnamefont{Buchinger}},
  \bibinfo{author}{\bibfnamefont{M.~T.} \bibnamefont{Burkey}},
  \bibnamefont{et~al.}, \bibinfo{journal}{Phys. Rev. Lett.}
  \textbf{\bibinfo{volume}{120}}, \bibinfo{pages}{262702}
  (\bibinfo{year}{2018}),
  \urlprefix\url{https://link.aps.org/doi/10.1103/PhysRevLett.120.262702}.

\bibitem[{\citenamefont{Lim and Holt}(2018)}]{yeunhwan2018}
\bibinfo{author}{\bibfnamefont{Y.}~\bibnamefont{Lim}} \bibnamefont{and}
  \bibinfo{author}{\bibfnamefont{J.~W.} \bibnamefont{Holt}},
  \bibinfo{journal}{Phys. Rev. Lett.} \textbf{\bibinfo{volume}{121}},
  \bibinfo{pages}{062701} (\bibinfo{year}{2018}),
  \urlprefix\url{https://link.aps.org/doi/10.1103/PhysRevLett.121.062701}.

\bibitem[{\citenamefont{Lovell and Nunes}(2018)}]{Lovell2018}
\bibinfo{author}{\bibfnamefont{A.~E.} \bibnamefont{Lovell}} \bibnamefont{and}
  \bibinfo{author}{\bibfnamefont{F.~M.} \bibnamefont{Nunes}},
  \bibinfo{journal}{Phys. Rev. C} \textbf{\bibinfo{volume}{97}},
  \bibinfo{pages}{064612} (\bibinfo{year}{2018}),
  \urlprefix\url{https://link.aps.org/doi/10.1103/PhysRevC.97.064612}.

\bibitem[{\citenamefont{Becchetti and Greenlees}(1969)}]{bg69}
\bibinfo{author}{\bibfnamefont{J.}~\bibnamefont{Becchetti},
  \bibfnamefont{F.D.}} \bibnamefont{and}
  \bibinfo{author}{\bibfnamefont{G.}~\bibnamefont{Greenlees}},
  \bibinfo{journal}{Phys. Rev.} \textbf{\bibinfo{volume}{182}},
  \bibinfo{pages}{1190} (\bibinfo{year}{1969}).

\bibitem[{\citenamefont{Varner et~al.}(1991)\citenamefont{Varner, Thompson,
  McAbee, Ludwig, and Clegg}}]{ch89}
\bibinfo{author}{\bibfnamefont{R.}~\bibnamefont{Varner}},
  \bibinfo{author}{\bibfnamefont{W.}~\bibnamefont{Thompson}},
  \bibinfo{author}{\bibfnamefont{T.}~\bibnamefont{McAbee}},
  \bibinfo{author}{\bibfnamefont{E.}~\bibnamefont{Ludwig}}, \bibnamefont{and}
  \bibinfo{author}{\bibfnamefont{T.}~\bibnamefont{Clegg}},
  \bibinfo{journal}{Phys. Rep.} \textbf{\bibinfo{volume}{201}},
  \bibinfo{pages}{57 } (\bibinfo{year}{1991}), ISSN \bibinfo{issn}{0370-1573}.

\bibitem[{\citenamefont{Koning and Delaroche}(2003)}]{kd2003}
\bibinfo{author}{\bibfnamefont{A.}~\bibnamefont{Koning}} \bibnamefont{and}
  \bibinfo{author}{\bibfnamefont{J.}~\bibnamefont{Delaroche}},
  \bibinfo{journal}{Nucl.Phys.} \textbf{\bibinfo{volume}{A713}},
  \bibinfo{pages}{231} (\bibinfo{year}{2003}).

\bibitem[{\citenamefont{Lovell et~al.}(2017)\citenamefont{Lovell, Nunes,
  Sarich, and Wild}}]{Lovell2017}
\bibinfo{author}{\bibfnamefont{A.~E.} \bibnamefont{Lovell}},
  \bibinfo{author}{\bibfnamefont{F.~M.} \bibnamefont{Nunes}},
  \bibinfo{author}{\bibfnamefont{J.}~\bibnamefont{Sarich}}, \bibnamefont{and}
  \bibinfo{author}{\bibfnamefont{S.~M.} \bibnamefont{Wild}},
  \bibinfo{journal}{Phys. Rev. C} \textbf{\bibinfo{volume}{95}},
  \bibinfo{pages}{024611} (\bibinfo{year}{2017}),
  \urlprefix\url{https://link.aps.org/doi/10.1103/PhysRevC.95.024611}.

\bibitem[{\citenamefont{King et~al.}(2018)\citenamefont{King, Lovell, and
  Nunes}}]{King2018}
\bibinfo{author}{\bibfnamefont{G.~B.} \bibnamefont{King}},
  \bibinfo{author}{\bibfnamefont{A.~E.} \bibnamefont{Lovell}},
  \bibnamefont{and} \bibinfo{author}{\bibfnamefont{F.~M.} \bibnamefont{Nunes}},
  \bibinfo{journal}{Phys. Rev. C} \textbf{\bibinfo{volume}{98}},
  \bibinfo{pages}{044623} (\bibinfo{year}{2018}),
  \urlprefix\url{https://link.aps.org/doi/10.1103/PhysRevC.98.044623}.

\bibitem[{\citenamefont{Thompson and Nunes}(2009)}]{ReactionsBook}
\bibinfo{author}{\bibfnamefont{I.~J.} \bibnamefont{Thompson}} \bibnamefont{and}
  \bibinfo{author}{\bibfnamefont{F.~M.} \bibnamefont{Nunes}},
  \emph{\bibinfo{title}{Nuclear Reactions for Astrophysics}}
  (\bibinfo{publisher}{Cambridge University Press}, \bibinfo{year}{2009}).

\bibitem[{\citenamefont{Lombardi et~al.}(1972)\citenamefont{Lombardi, Boyd,
  Arking, and Robbins}}]{48cap14}
\bibinfo{author}{\bibfnamefont{J.}~\bibnamefont{Lombardi}},
  \bibinfo{author}{\bibfnamefont{R.}~\bibnamefont{Boyd}},
  \bibinfo{author}{\bibfnamefont{R.}~\bibnamefont{Arking}}, \bibnamefont{and}
  \bibinfo{author}{\bibfnamefont{A.}~\bibnamefont{Robbins}},
  \bibinfo{journal}{Nuclear Physics A} \textbf{\bibinfo{volume}{188}},
  \bibinfo{pages}{103 } (\bibinfo{year}{1972}), ISSN \bibinfo{issn}{0375-9474},
  \urlprefix\url{http://www.sciencedirect.com/science/article/pii/0375947472901868}.

\bibitem[{\citenamefont{Mueller et~al.}(2011)\citenamefont{Mueller, Charity,
  Shane, Sobotka, Waldecker, Dickhoff, Crowell, Esterline, Fallin, Howell
  et~al.}}]{48can12}
\bibinfo{author}{\bibfnamefont{J.~M.} \bibnamefont{Mueller}},
  \bibinfo{author}{\bibfnamefont{R.~J.} \bibnamefont{Charity}},
  \bibinfo{author}{\bibfnamefont{R.}~\bibnamefont{Shane}},
  \bibinfo{author}{\bibfnamefont{L.~G.} \bibnamefont{Sobotka}},
  \bibinfo{author}{\bibfnamefont{S.~J.} \bibnamefont{Waldecker}},
  \bibinfo{author}{\bibfnamefont{W.~H.} \bibnamefont{Dickhoff}},
  \bibinfo{author}{\bibfnamefont{A.~S.} \bibnamefont{Crowell}},
  \bibinfo{author}{\bibfnamefont{J.~H.} \bibnamefont{Esterline}},
  \bibinfo{author}{\bibfnamefont{B.}~\bibnamefont{Fallin}},
  \bibinfo{author}{\bibfnamefont{C.~R.} \bibnamefont{Howell}},
  \bibnamefont{et~al.}, \bibinfo{journal}{Phys. Rev. C}
  \textbf{\bibinfo{volume}{83}}, \bibinfo{pages}{064605}
  (\bibinfo{year}{2011}),
  \urlprefix\url{https://link.aps.org/doi/10.1103/PhysRevC.83.064605}.

\bibitem[{\citenamefont{McCamis et~al.}(1986)\citenamefont{McCamis, Nasr,
  Birchall, Davison, van Oers, Verheijen, Carlson, Cox, Clark, Cooper
  et~al.}}]{48cap25}
\bibinfo{author}{\bibfnamefont{R.~H.} \bibnamefont{McCamis}},
  \bibinfo{author}{\bibfnamefont{T.~N.} \bibnamefont{Nasr}},
  \bibinfo{author}{\bibfnamefont{J.}~\bibnamefont{Birchall}},
  \bibinfo{author}{\bibfnamefont{N.~E.} \bibnamefont{Davison}},
  \bibinfo{author}{\bibfnamefont{W.~T.~H.} \bibnamefont{van Oers}},
  \bibinfo{author}{\bibfnamefont{P.~J.~T.} \bibnamefont{Verheijen}},
  \bibinfo{author}{\bibfnamefont{R.~F.} \bibnamefont{Carlson}},
  \bibinfo{author}{\bibfnamefont{A.~J.} \bibnamefont{Cox}},
  \bibinfo{author}{\bibfnamefont{B.~C.} \bibnamefont{Clark}},
  \bibinfo{author}{\bibfnamefont{E.~D.} \bibnamefont{Cooper}},
  \bibnamefont{et~al.}, \bibinfo{journal}{Phys. Rev. C}
  \textbf{\bibinfo{volume}{33}}, \bibinfo{pages}{1624} (\bibinfo{year}{1986}),
  \urlprefix\url{https://link.aps.org/doi/10.1103/PhysRevC.33.1624}.

\bibitem[{\citenamefont{Dickens et~al.}(1968)\citenamefont{Dickens, Eichler,
  and Satchler}}]{90zrp12}
\bibinfo{author}{\bibfnamefont{J.~K.} \bibnamefont{Dickens}},
  \bibinfo{author}{\bibfnamefont{E.}~\bibnamefont{Eichler}}, \bibnamefont{and}
  \bibinfo{author}{\bibfnamefont{G.~R.} \bibnamefont{Satchler}},
  \bibinfo{journal}{Phys. Rev.} \textbf{\bibinfo{volume}{168}},
  \bibinfo{pages}{1355} (\bibinfo{year}{1968}),
  \urlprefix\url{https://link.aps.org/doi/10.1103/PhysRev.168.1355}.

\bibitem[{\citenamefont{Wang and Rapaport}(1990)}]{90zrn10}
\bibinfo{author}{\bibfnamefont{Y.}~\bibnamefont{Wang}} \bibnamefont{and}
  \bibinfo{author}{\bibfnamefont{J.}~\bibnamefont{Rapaport}},
  \bibinfo{journal}{Nuclear Physics A} \textbf{\bibinfo{volume}{517}},
  \bibinfo{pages}{301 } (\bibinfo{year}{1990}), ISSN \bibinfo{issn}{0375-9474},
  \urlprefix\url{http://www.sciencedirect.com/science/article/pii/037594749090037M}.

\bibitem[{\citenamefont{Ball et~al.}(1964)\citenamefont{Ball, Fulmer, and
  Bassel}}]{90zrp22}
\bibinfo{author}{\bibfnamefont{J.~B.} \bibnamefont{Ball}},
  \bibinfo{author}{\bibfnamefont{C.~B.} \bibnamefont{Fulmer}},
  \bibnamefont{and} \bibinfo{author}{\bibfnamefont{R.~H.}
  \bibnamefont{Bassel}}, \bibinfo{journal}{Phys. Rev.}
  \textbf{\bibinfo{volume}{135}}, \bibinfo{pages}{B706} (\bibinfo{year}{1964}),
  \urlprefix\url{https://link.aps.org/doi/10.1103/PhysRev.135.B706}.

\bibitem[{\citenamefont{Makofske et~al.}(1972)\citenamefont{Makofske,
  Greenlees, Liers, and Pyle}}]{208pbp16}
\bibinfo{author}{\bibfnamefont{W.}~\bibnamefont{Makofske}},
  \bibinfo{author}{\bibfnamefont{G.~W.} \bibnamefont{Greenlees}},
  \bibinfo{author}{\bibfnamefont{H.~S.} \bibnamefont{Liers}}, \bibnamefont{and}
  \bibinfo{author}{\bibfnamefont{G.~J.} \bibnamefont{Pyle}},
  \bibinfo{journal}{Phys. Rev. C} \textbf{\bibinfo{volume}{5}},
  \bibinfo{pages}{780} (\bibinfo{year}{1972}),
  \urlprefix\url{https://link.aps.org/doi/10.1103/PhysRevC.5.780}.

\bibitem[{\citenamefont{{Floyd}}(1981)}]{208pbn16}
\bibinfo{author}{\bibfnamefont{C.~E.} \bibnamefont{{Floyd}},
  \bibfnamefont{Jr.}}, Ph.D. thesis, \bibinfo{school}{DUKE UNIVERSITY.}
  (\bibinfo{year}{1981}).

\bibitem[{\citenamefont{van Oers et~al.}(1974)\citenamefont{van Oers, Haw,
  Davison, Ingemarsson, Fagerstr\"om, and Tibell}}]{208pbp35}
\bibinfo{author}{\bibfnamefont{W.~T.~H.} \bibnamefont{van Oers}},
  \bibinfo{author}{\bibfnamefont{H.}~\bibnamefont{Haw}},
  \bibinfo{author}{\bibfnamefont{N.~E.} \bibnamefont{Davison}},
  \bibinfo{author}{\bibfnamefont{A.}~\bibnamefont{Ingemarsson}},
  \bibinfo{author}{\bibfnamefont{B.}~\bibnamefont{Fagerstr\"om}},
  \bibnamefont{and} \bibinfo{author}{\bibfnamefont{G.}~\bibnamefont{Tibell}},
  \bibinfo{journal}{Phys. Rev. C} \textbf{\bibinfo{volume}{10}},
  \bibinfo{pages}{307} (\bibinfo{year}{1974}),
  \urlprefix\url{https://link.aps.org/doi/10.1103/PhysRevC.10.307}.

\bibitem[{\citenamefont{Thompson}(1988)}]{fresco}
\bibinfo{author}{\bibfnamefont{I.~J.} \bibnamefont{Thompson}},
  \bibinfo{journal}{Comput. Phys. Rept.} \textbf{\bibinfo{volume}{7}},
  \bibinfo{pages}{167} (\bibinfo{year}{1988}).

\bibitem[{\citenamefont{Johnson and Tandy}(1974)}]{Johnson1974}
\bibinfo{author}{\bibfnamefont{R.}~\bibnamefont{Johnson}} \bibnamefont{and}
  \bibinfo{author}{\bibfnamefont{P.}~\bibnamefont{Tandy}},
  \bibinfo{journal}{Nuclear Physics A} \textbf{\bibinfo{volume}{235}},
  \bibinfo{pages}{56 } (\bibinfo{year}{1974}), ISSN \bibinfo{issn}{0375-9474},
  \urlprefix\url{http://www.sciencedirect.com/science/article/pii/037594747490178X}.

\bibitem[{\citenamefont{Titus et~al.}(2016)\citenamefont{Titus, Ross, and
  Nunes}}]{nlat}
\bibinfo{author}{\bibfnamefont{L.}~\bibnamefont{Titus}},
  \bibinfo{author}{\bibfnamefont{A.}~\bibnamefont{Ross}}, \bibnamefont{and}
  \bibinfo{author}{\bibfnamefont{F.}~\bibnamefont{Nunes}},
  \bibinfo{journal}{Comput. Phys. Commun.} \textbf{\bibinfo{volume}{207}},
  \bibinfo{pages}{499 } (\bibinfo{year}{2016}), ISSN \bibinfo{issn}{0010-4655},
  \urlprefix\url{http://www.sciencedirect.com/science/article/pii/S0010465516302028}.

\bibitem[{\citenamefont{Metz et~al.}(1975)\citenamefont{Metz, Callender, and
  Bockelman}}]{ca48dp}
\bibinfo{author}{\bibfnamefont{W.~D.} \bibnamefont{Metz}},
  \bibinfo{author}{\bibfnamefont{W.~D.} \bibnamefont{Callender}},
  \bibnamefont{and} \bibinfo{author}{\bibfnamefont{C.~K.}
  \bibnamefont{Bockelman}}, \bibinfo{journal}{Phys. Rev. C}
  \textbf{\bibinfo{volume}{12}}, \bibinfo{pages}{827} (\bibinfo{year}{1975}),
  \urlprefix\url{https://link.aps.org/doi/10.1103/PhysRevC.12.827}.

\end{thebibliography}

\end{document}